# Exploring electric field sensing for solid-state nanopores

*Muhammad Sajeer P, Manoj M.Varma*
*Centre for Nanoscience and Engineering, Indian Institute of Science, Bangalore, 560012*
*Corresponding email:muhammads@iisc.ac.in*

Abstract
Solid-state nanopores have received substantial attention in the past years owing to their simplicity and potential applications expected in genomics, sensing, archival information storage, and computing. The underlying sensing technique of nanopore technology is the analysis of modulations in the ionic current while molecules are electrophoretically driven through the nanopore. This current blockade-based sensing is presently well recognized and commercially used for applications such as DNA sequencing. However, this ionic current-based method has limitations and increased complexity for futuristic applications such as single molecular protein sequencing, where diverse charges and shape distributions are involved. A high throughput readout method that can be used in extreme environments and has improved sensitivity to the mixed charge profiles and shape of the analytes is required. In this work, we present an exploratory finite element simulation study on the feasibility of using electric-field modulations instead of ionic current blockades for nanopore translocation measurements. This electric field sensing technique has further advantages over ionic current blockade measurements. For instance, Electric field sensing is capable of size and charge discretion with lesser noise and does not mandate the presence of an electrolyte solution. This technique can be used in extreme environments and developed for defense and space applications such as detecting air-born particles and future mars, moon, and Europa missions. We hope this work will be a starting point for developing electric field sensing for nanopore applications and opening the field of nanopore electrometry.

1. Introduction

Solid-state nanopores have received significant attention in the past years owing to their simplicity, relatively lower cost, and potential applications in genomics[1], single-molecule sensing[2], DNA-based information storage[3], and computing[4]. The solid-state nanopores also have substantial advantages, such as chemical, mechanical, and thermal stability and easy integration with semiconductor processes, over their biological counterparts. In the nanopore translocation measurement setup, the nanopore is situated between two chambers containing an electrolyte, typically potassium chloride (KCl) solution. The target molecules are electrophoretically driven from one chamber to the other through the nanopore. The movement of target molecules through the pore, referred to as translocation, causes transient blockades of the ionic current through the nanopore, which is recorded. These signals can be deciphered to correlate with the charge, size, and concentration of target molecules. This current blockade-based measurement techniques are now well-established and commercially used for DNA sequencing using biological nanopores. Here, features in the raw ionic current traces are used to map them to one of the four canonical bases using machine learning-based approaches[5]. One can extend this mapping to include modified bases as well, but even with their inclusion, the number of distinct entities that the current must be mapped onto is of the order of five or six. On the other hand, emerging applications such as single-molecule protein sequencing require mapping the nanopore current blockades to 20 amino acids. Also, proteins exhibit orders of magnitude higher sequence diversity than DNA or RNA. For instance, mammalian cells have about 30,000 fold more protein molecules than mRNA[6]. Single-molecule protein sequencing thus presents significant challenges compared to DNA sequencing, as discussed in the perspective article by Savlov et al[6]. Thus, aside from the problem of mapping current signals to a larger number of independent entities, more challenging throughput requirements are also there to decode the abundant proteins. Other applications, such as DNA based information storage, may also require considerably faster readout than the ~ 500 bases/sec speed offered by the current blockade-based readout. For instance, a readout method that is sensitive to inhomogeneities in charge distributions and geometry, as opposed to current blockades, which are predominantly determined by the total volume and charge, may provide a more effective method to handle the complexities presented by single-molecule protein sequencing.

With these considerations in mind, we explored the feasibility of using electric-field modulations instead of ionic current blockades to measure nanopore translocations. Electric field-based sensing has several potential advantages. The local electric field is inherently sensitive to the charge distribution of the target molecule and, therefore, potentially encodes richer information about the molecule. Another aspect of electric-field sensing is that it can be done without using an electrolyte, even in vacuum. Thus, this approach may have potential benefits in the direct monitoring of air-borne particles as well as in extreme environments such as outer space or hot and freezing conditions. Electric field sensing is presently widely used in macroscale in proximity and object identifier sensors[7–10]. On the other hand, nanopore signal readout using electric-field modulations will require highly localized measurements. We expect that the recent development of nano[11] and micro-scale[12] electric field sensors, nanoscale electric field imaging using nitrogen-vacancy centere[13] and single spins[14], and the development of electric field imaging technology using polarized neutrones[15] can possibly be modified for electric field measurements in solid-state nanopores. In this study, we performed finite element simulations of translocations of a particle through an hourglass-shaped silicon nitride nanopore (Fig 1A). We used these simulations to compare the ionic current and local electric field modulations caused by the translocating particle under various conditions, such as the size of the particle and magnitude and polarity of charge, as described in the subsequent sections. We hope this work will be a starting point of for developing electric field sensing techniques for exciting nanopore-related applications and opening the field of nanopore electrometry.

2. Methods

The finite element simulation is conducted by solving the Poisson-Nernst-Plank (PNP) equation in COMSOL (Version 5.5) software. The Poisson equation provides information on the electric potential distribution, and Nernst-Plank equation solves for the diffusion of ions under external voltage bias[16] (refer to supplementary information). An hourglass-shaped silicon nitride nanopore is defined using the 2D axisymmetric model in COMSOL. The nanopore defined here has a length of 10 nm, an opening diameter of 5nm, and a centre constriction of 3nm (Figure 1A). The particle, which is represented by a sphere, is translocated through the center of the pore along the Z-axis to avoid off-axis effects[16]. The translocation is done step by step in a time-independent manner. A voltage bias of 1V is applied across the pore, and figure 1(b-c) shows the simulated voltage distribution in the nanopore system. Figure 1(d) indicates that the voltage drop's significant share occurs near the pore region. The details of the simulation, such as the parameters, geometry, and reliability, are discussed in the supplementary information.

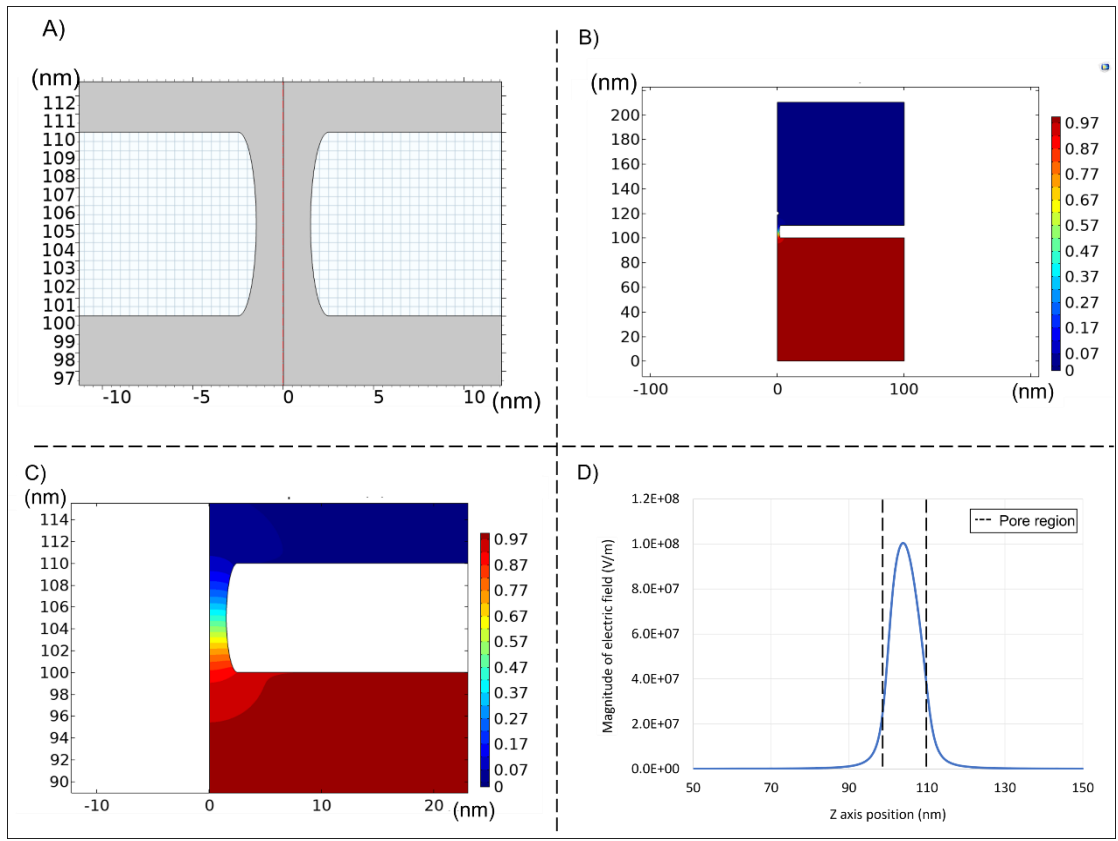

Figure 1] *A. Hourglass-shaped solid-state nanopore geometry created in COMSOL. B) The voltage distribution in the complete nanopore system represented using an axisymmetric model in COMSOL C) Magnified image of an area near the nanopore showing voltage distribution D) Variation in magnitude of the electric field along the line through the center of the pore (z-axis). The dotted line shows the start and end of the nanopore. The maximum electric field is focused near the pore and reduces as we move away from the pore as maximum potential drop happens near the pore due to the larger resistance of the pore compared to the electrolyte solution*

3. Results

    3.1 The electric field is sensitive to the size of translocating particle

The size variations of the translocating molecule can cause modulations in current blockade due to the volume exclusion of ions from the bulk of the pore. The scaling of contrast in ionic current blockade for variation in the particle size (diameter of 0.5nm to 2.5nm) is shown in figure 2B. To explore if the electric field has a similar discretion capability, the electric field sensor is placed at the center and edges of the nanopore (figure 2A inset). The changes in the magnitude of the electric field are calculated for translocating particles having varying sizes (diameter from 0.5nm to 2.5nm), as shown in figure 2B. This particle has a surface charge of 50mC/m$^2$. This simulation indicates that the contrast in the magnitude of an electric field can be mapped to variation in the translocating particle's size. This can potentially be used for size discretion of translocating molecules in single molecular sensing applications. It is important to remember that the total electric field magnitude measured by the sensor will be a cumulative value arising from various sources, such as the charge of the analyte, the surface charge of nanopores, and ions.

We have normalized the current and electric field values by dividing them by the base value to compare ionic current and electric field sensing. Then signal contrast [maximum value-minimum value] is calculated for the particle of 2.5nm diameter translocating through the hourglass-shaped pore. Figure 2A shows the comparison between the ionic current and the electric field. It is important to note that the particle is translocating in the +Z axis direction, starting from 90 nm to 120 nm, with a step size of 0.25nm. This analysis shows that the electric field is comparable with current blockade and sensitive to the size of the particle that translocates through the pore. Hence, electric field sensing can potentially be used for nanopore sensing applications as an alternative or complementary method to the existing current blockade-based technique.

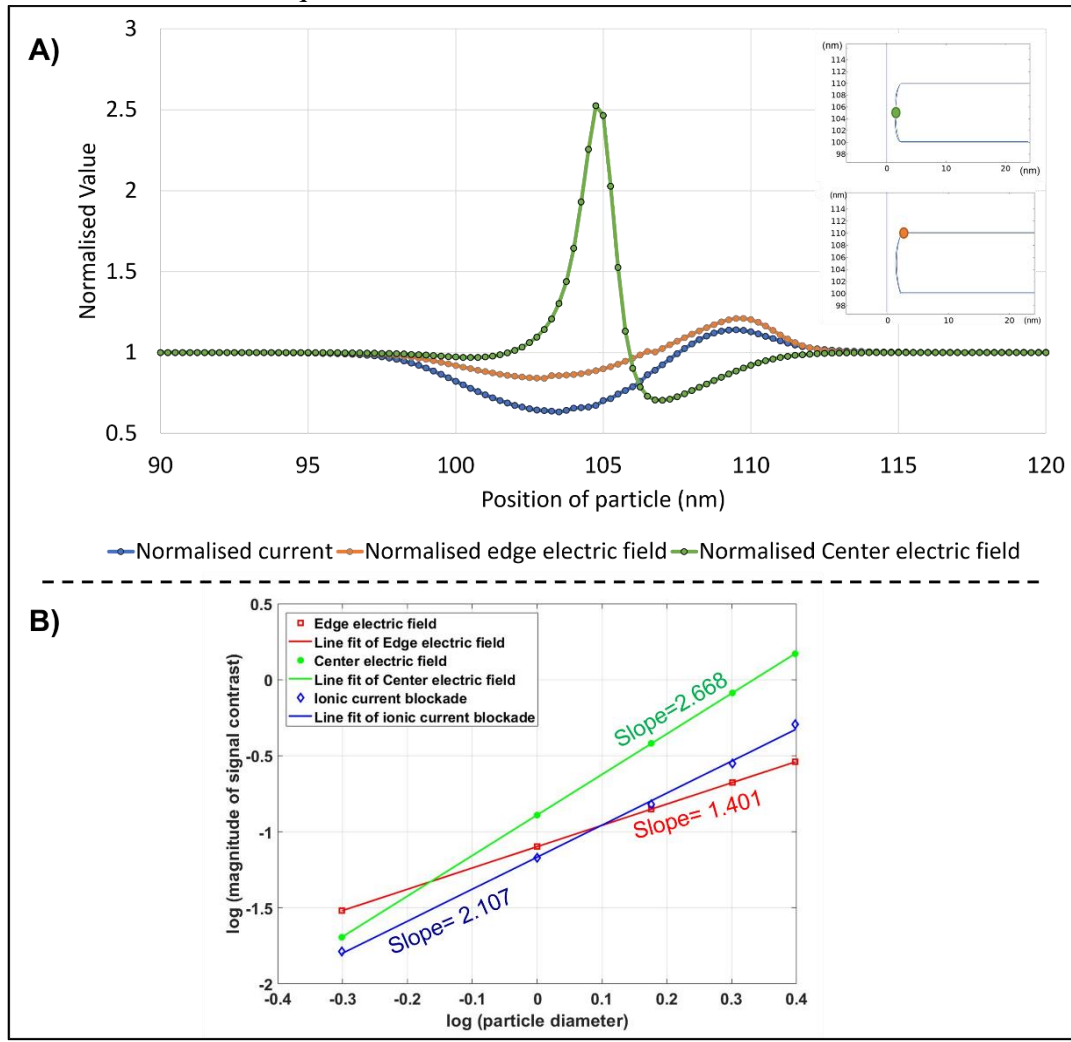

Figure 2: *A) Comparison between normalized ionic current blockade signal (blue) and electric field from edge sensor (green dot in inset) and center sensor (orange dot in inset). B) The scaling of the Electric field and ionic current blockade with respect to the varying particle size. The slope is 1.401 (0.8541, 1.949) for the edge electric field, 2.668 (2.238, 3.099) for the center electric field, and 2.107 (1.94, 2.26) for ionic current blockade obtained from linear curve fitting in MATLAB.*

### 3.2 Feasibility of charge discrimination using electric field sensing

The existing ionic current blockade method using solid-state nanopores can sense the charge variation of particles mainly through techniques such as surface functionalization of nanopores[17]. Hence, there is a need for less complex methods capable of direct sensing the magnitude and polarity of the charge of particles. As the surface charge of molecules predominantly contributes to its biological process[18], it is also crucial to develop methods to study the charge profile of molecules. This Multiphysics study, for instance, by taking examples of uniformly charged molecules, shows that electric field sensing can characterize particles based on their surface charge.

Here, the $\frac{\Delta I}{I}$ and $\frac{\Delta E}{E}$ has been calculated for the particle's surface charge ranging from -500 to 500 mC/m$^2$ . The $\Delta I$ is the amplitude of the current blockade [maximum ionic current- minimum ionic current], and $\Delta E$ is the contrast [maximum value – minimum value] of the electric field. I and E are the base values of the current and electric field when the particle is not translocating through the pore. This charge range is chosen considering the reported values in literature, such as the surface charge of B-DNA: -163 mC/m$^2$ [19], dsDNA: -150mC/m$^2$, Streptavidin: -17.1mC/m$^2$ and DNA origami structures: -12.9mC/m$^2$ [20]. Figure 3 shows that electric field sensing is competent in differentiating particles based on the charge and their polarity compared to ionic current blockade sensing. It has been observed that for particles having -125mC/m$^2$ < charge < ~275mC/m$^2$, the center Electric field sensor is the most effective way to differentiate them as $\Delta E$ is better than the current blockade in this bracket. The current blockade is the better candidate for detecting higher charge densities. We suspect that ions of opposite charges accumulate on the particle on higher charges, which increases the effective volume and masks the charge of the particle. The contour map of variation in the electric field during particle translocation is also plotted (figure 3 of supplementary information). In the case of the neutral particle, the signal contrast for the neutral particle is 0.186 for electric field measurement from the sensor situated at the edge of the pore and 0.203 for the current blockade, which is nearly equivalent in magnitude. In the case of the electric field sensor at the centre of the pore (figure 2a), the electric field's contrast is 1.37, which is ~ 7 times better than the contrast of the current blockade signal. This Multiphysics study shows that the electric field has a similar capability, or better in some cases, compared to current blockade measurements for single molecular sensing applications.

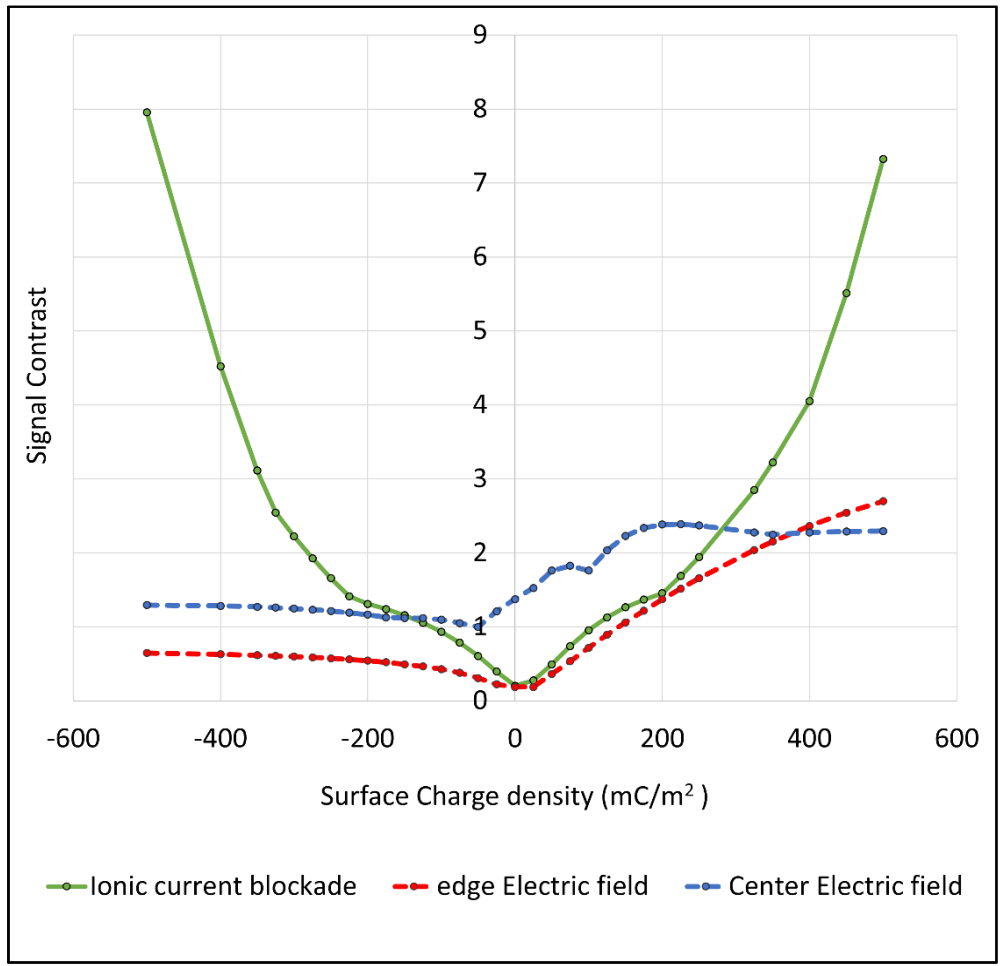

Figure 3) *Comparison between Ionic current blockade and modulations in the electric field for the particle of varying surface charges. The signal contrast is the maximum – minimum value during translocation through the nanopore.*

### 3.3 Better precision in translocation measurement using electric field sensing

Connecting a single entity of ionic current measurement to all translocation events and resolving the same is a challenging task, especially in cases where analyzing body fluids such as serum, where random translocations will be involved[21]. In contrast to ionic current blockade-based measurements, multiple electric field sensors can be used to simultaneously measure a single translocation event. For instance, electric field sensors can be placed in multiple places nearby the nanopore. As the electric field signal from different EF sensors has different electric field modulations (Figure 4), this can potentially reduce the error rate statistically. This aspect is vital during the live screening of particles for defense applications. This will also help in an easy understanding of the random translocation happening in body fluids as now we are equipped with multiple hands of electric field and ionic current measurements for the analysis. Unlike in the lab, where multiple copies of DNAs or specific analytes are available, it may not always be the case in the field where the supply of analytes is limited, or the situation may not allow multiple measurements. Hence, developing the electric field-based sensing method for nanopore technology is important, which can either be combined with the existing ionic current blockade measurement as a complementary technique or can be developed independently.

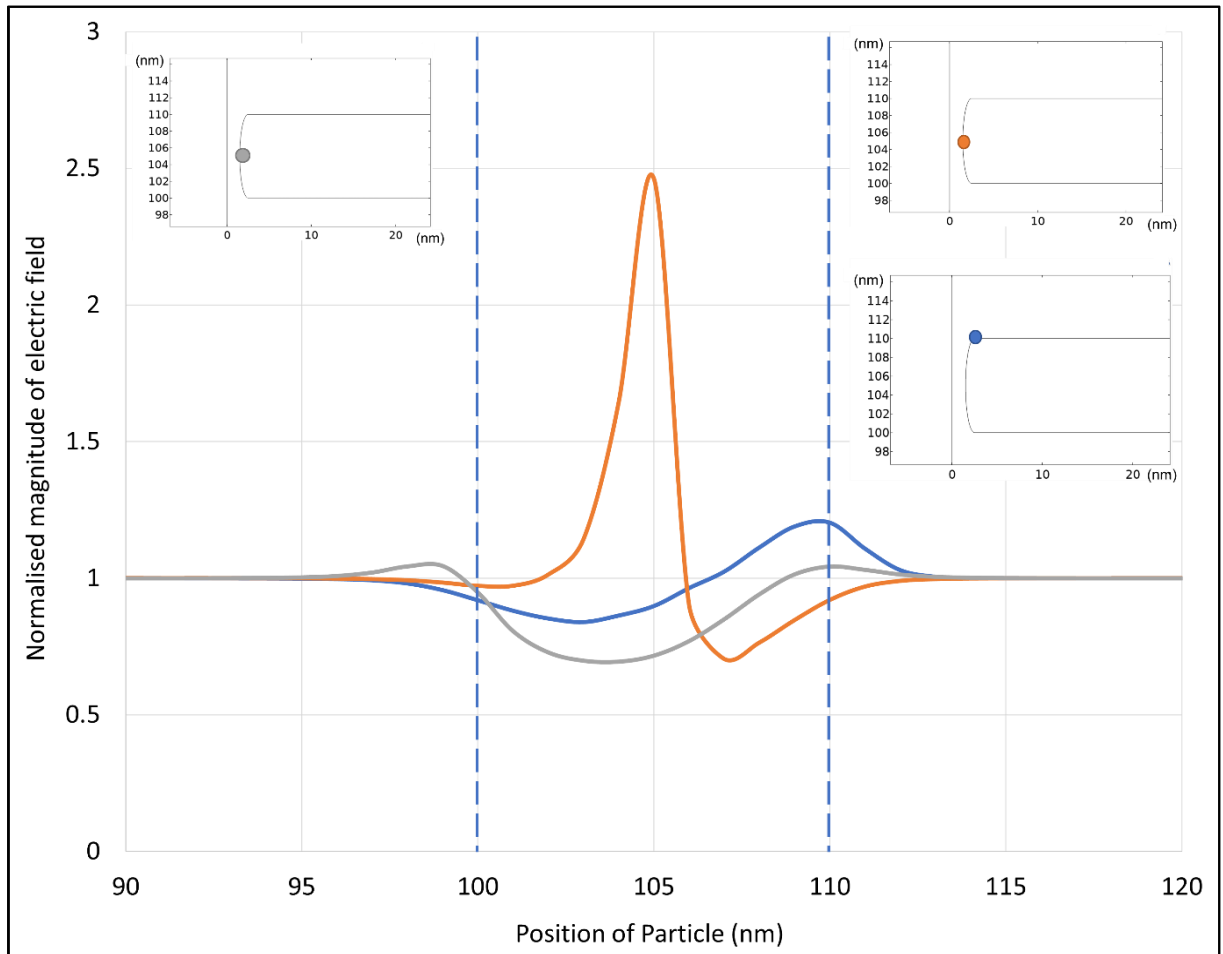

Figure 4: *Localized E-fields have different responses that can aid in error correction. The modulation in electric field during the translocation of 50mC/m$^2$ with respect to sensors positioned at various places near the pore (in inlet) shown in the figure. Colors of sensors and electric field are matched for easy identification.*

## 4  Conclusion

Solid-state nanopores are gaining industrial and academic interest due to their business potential and various applications ranging from genetics to filtering and their robustness and durability compared to biological pores. Unlike nanopore based DNA sequencing, correlating ionic current blockades to twenty amino acids raises significant complexity. There are also high throughput requirements. In this work, we have presented an exploratory Multiphysics simulation study on the feasibility of using electric-field modulations instead of ionic current blockades to measure nanopore translocations.

1. The Multiphysics simulation shows that electric field sensing is comparable with the sensing capabilities of the existing ionic current-blockade based techniques, where electric field sensing gives better signal contrast in some cases.
2. Electric field sensing also distinguishes particles based on charge magnitude and polarity.
3. Placing multiple electric field sensors on the nanopore can lead to multiple simultaneous measurements of a single transportation event of analytes through the pore, potentially reducing the error rate and saving time. For instance, this opens the possibility for faster DNA and protein sequencing compared to existing methods.

4. As the ionic solution is not mandatory for electric field-based sensing techniques, it has potential benefits in direct monitoring of air-borne particles as well as in extreme environments such as outer space or hot and freezing conditions
5. Unlike ionic current blockade, this method, based on the detection of an electrical property, the electric field, can be easily incorporated into digital data storage applications.
6. Electric field sensing equips us with the ability to track particle dynamics inside nanopores and can possibly use for the discretion of particle shapes.

This work aimed to check the capability of electric field sensing as an alternative or complementary technique to current blockade measurements, and the same has been justified. Even though these are yet to be verified experimentally, we hope this work will open many possibilities for electric field measurement-based nanopore sensing and extend the capability and applications of solid-state nanopores. Welcome to the era of nano electrometry-based nanopore sensing.

## Acknowledgment

We thank Sohini Pal (Postdoc, University of California San Diego, US) for COMSOL related discussions. Thanks to Tuhin Chakraborty (Postdoc, Georgia Tech) for the computational infrastructure support. Thanks to Prof. SVM Satyanarayana, Pondicherry University and Prathyay Sarkar (BSc Research, IISc Bangalore) for the discussions regarding the electric field. We also acknowledge other nanopore group (https://sites.google.com/view/nanoporegroup/home) Members, namely Anumol Dominic, Anu Roshni, Avisekh Pal, Mayank Mitram, Pranjal Sur, Simran, Shreesumanas, Jiase Johnson and Muddukrishna for providing their critical suggestions during discussions and thanks for being the wonderful people who motivate us, laugh with us and keeping up our research spirit and mental well-being. We also thank the Centre for Nanoscience and Engineering at the Indian Institute of Science, Bengaluru, for providing a supportive environment. This work was supported by the Scientific and Useful Profound Research Advancement (SUPRA) Program of the Science Engineering Research Board (SERB) under Grant SPR/2021/000275.

## Author Declarations

Conflicts of interests: Authors have no conflicts to disclose

Author Contributions: MSP and MV have conceptualized the idea. MSP conducted the simulations with suggestions from MV and analysed the data. MV acquired funding and supervised this study. MSP wrote the initial manuscript, and MV did the review and editing.

Data Availability: Data available on request from the authors

## References

1. Majd, S. *et al.* Applications of biological pores in nanomedicine, sensing, and

   nanoelectronics. *Curr. Opin. Biotechnol.* **21**, 439–476 (2010).

2. Gu, L.-Q. & Shim, J. W. Single molecule sensing by nanopores and nanopore devices. *The*

   *Analyst* **135**, 441–451 (2010).

3. Chen, K. *et al.* Digital Data Storage Using DNA Nanostructures and Solid-State Nanopores. *Nano Lett.* **19**, 1210–1215 (2019).

4. Zhang, K. *et al.* A nanopore interface for higher bandwidth DNA computing. *Nat. Commun.* **13**, 4904 (2022).

5. Wan, Y. K., Hendra, C., Pratanwanich, P. N. & Göke, J. Beyond sequencing: machine learning algorithms extract biology hidden in Nanopore signal data. *Trends Genet.* **38**, 246–257 (2022).

6. MacCoss, M. J., Alfaro, J., Wanunu, M., Faivre, D. A. & Slavov, N. Sampling the proteome by emerging single-molecule and mass-spectrometry methods. (2022) doi:10.48550/ARXIV.2208.00530.

7. Wilmsdorff, J. von, Kirchbuchner, F., Fu, B., Braun, A. & Kuijper, A. An experimental overview on electric field sensing. *J. Ambient Intell. Humaniz. Comput.* **10**, 813–824 (2019).

8. Bian, Sizhen. Human Activity Recognition with Field Sensing Technique. 37516 KB, VII, 202 pages (Technische Universität Kaiserslautern, 2022). doi:10.26204/KLUEDO/6922.

9. Noras, M. A. Activity Detection and Recognition With Passive Electric Field Sensors. *IEEE Trans. Ind. Appl.* **58**, 800–806 (2022).

10. Tang, X. & Mandal, S. Indoor Occupancy Awareness and Localization Using Passive Electric Field Sensing. *IEEE Trans. Instrum. Meas.* **68**, 4535–4549 (2019).

11. Rajasekar, R. & Robinson, S. Nano-electric field sensor based on Two Dimensional Photonic Crystal resonator. *Opt. Mater.* **85**, 474–482 (2018).

12. Han, Z., Xue, F., Hu, J. & He, J. Micro Electric Field Sensors: Principles and Applications. *IEEE Ind. Electron. Mag.* **15**, 35–42 (2021).

13. Bian, K. *et al.* Nanoscale electric-field imaging based on a quantum sensor and its charge-state control under ambient condition. *Nat. Commun.* **12**, 2457 (2021).

14. Dolde, F. *et al.* Electric-field sensing using single diamond spins. *Nat. Phys.* **7**, 459–463 (2011).

15. Jau, Y.-Y., Hussey, D. S., Gentile, T. R. & Chen, W. Electric Field Imaging Using Polarized Neutrons. *Phys. Rev. Lett.* **125**, 110801 (2020).

16. Ying, C., Houghtaling, J. & Mayer, M. Effects of off-axis translocation through nanopores on the determination of shape and volume estimates for individual particles. *Nanotechnology* **33**, 275501 (2022).

17. Chen, S. *et al.* A glass nanopore ionic sensor for surface charge analysis. *RSC Adv.* **10**, 21615–21620 (2020).

18. Lin, C.-Y. *et al.* Modulation of Charge Density and Charge Polarity of Nanopore Wall by Salt Gradient and Voltage. *ACS Nano* **13**, 9868–9879 (2019).

19. Kominami, H., Kobayashi, K. & Yamada, H. Molecular-scale visualization and surface charge density measurement of Z-DNA in aqueous solution. *Sci. Rep.* **9**, 6851 (2019).

20. Yamamoto, Y., Kominami, H., Kobayashi, K. & Yamada, H. Surface charge density measurement of a single protein molecule with a controlled orientation by AFM. *Biophys. J.* **120**, 2490–2497 (2021).

21. Shen, B. *et al.* Advanced DNA Nanopore Technologies. *ACS Appl. Bio Mater.* **3**, 5606–5619 (2020).